\documentclass[sigconf, authorversion, nonacm]{acmart}
\newcommand{\R}{\mathbb{R}}

\AtBeginDocument{%
  \providecommand\BibTeX{{%
    \normalfont B\kern-0.5em{\scshape i\kern-0.25em b}\kern-0.8em\TeX}}}

\begin{document}

% \title[Intelligent Trading Systems: A Sentiment-Aware Reinforcement Learning Approach]{Intelligent Trading Systems: A Sentiment-Aware Reinforcement Learning Approach}
\title{Intelligent Trading Systems: A Sentiment-Aware Reinforcement Learning Approach}

\author{Francisco Caio Lima Paiva}
\email{francisco.paiva@usp.br}
\orcid{0000-0003-2453-4608}
\affiliation{%
  \institution{Universidade de São Paulo}
  \city{São Paulo}
  \state{SP}
  \country{Brazil}
}

\author{Leonardo Kanashiro Felizardo}
\email{leonardo.felizardo@usp.br}
\orcid{0000-0002-2871-860X}
\affiliation{%
  \institution{Universidade de São Paulo}
  \city{São Paulo}
  \state{SP}
  \country{Brazil}
}

\author{Reinaldo Augusto da Costa Bianchi}
\email{rbianchi@fei.edu.br}
\orcid{0000-0001-9097-827X}
\affiliation{%
  \institution{Centro Universitário FEI}
  \streetaddress{Av. Humberto de Alencar Castelo Branco, 3972-B}
  \city{São Bernardo do Campo}
  \state{SP}
  \country{Brazil}
  \postcode{09850-901}
}

\author{Anna Helena Reali Costa}
\email{anna.reali@usp.br}
\orcid{0000-0001-7309-4528}
\affiliation{%
  \institution{Universidade de São Paulo}
  \city{São Paulo}
  \state{SP}
  \country{Brazil}
}

\renewcommand{\shortauthors}{Lima Paiva, et al.}

\begin{abstract}

The feasibility of making profitable trades on a single asset on stock exchanges based on patterns identification has long attracted researchers. Reinforcement Learning (RL) and Natural Language Processing have gained notoriety in these single-asset trading tasks, but only a few works have explored their combination. Moreover, some issues are still not addressed, such as extracting market sentiment momentum through the explicit capture of sentiment features that reflect the market condition over time and assessing the consistency and stability of RL results in different situations. Filling this gap, we propose the Sentiment-Aware RL (SentARL) intelligent trading system that improves profit stability by leveraging market mood through an adaptive amount of past sentiment features drawn from textual news. We evaluated SentARL across twenty assets, two transaction costs, and five different periods and initializations to show its consistent effectiveness against baselines. Subsequently, this thorough assessment allowed us to identify the boundary between news coverage and market sentiment regarding the correlation of price-time series above which SentARL's effectiveness is outstanding.

\end{abstract}





% \keywords{datasets, neural networks, gaze detection, text tagging}

\maketitle

\section{Introduction}
\label{sec:intro}

Investing in the financial market is an increasingly common activity in which a person can select assets (e.g., stocks, currencies, and others) to invest for a given period. 
In financial markets, \textit{active trading} is the procedure in which investors make short-term moves by changing their asset positions (i.e., long, neutral, or short) in order to profit from price movements and following some custom-made strategy.
Usually, active trading encompasses observing past asset information, such as price and volume, to search for patterns that might indicate future trends. However, under the optics of the principles studied by the economics research community, there is an open discussion regarding the predictability of stock markets. 
Some argue in favor of the \textit{Efficient Market Hypothesis} (EMH), which states that exploitability is not possible in an efficient market because the price of an asset adjusts almost instantly to the impact of events. 
On the other hand, the \textit{Adaptive Market Hypothesis} (AMH)~\cite{Lo2004} defines different levels of market efficiency related to the liquidity of an asset. 
Thus, according to AMH, depending on the asset or venue, it may be possible to exploit market inefficiencies.

Given this challenging, complex, and repetitive nature of active trading, the \textit{machine learning} (ML) community has naturally taken an interest in intelligent trading systems. 
For instance, a systematic review of ML works \cite{Henrique2019} identified the prevalence of approaches that look at financial trading as a market prediction problem and thus resort to supervised learning techniques. 
However, although prediction based on supervised learning has been successful, the introduction of typical trading operating costs may make it less effective than a formulation based on sequential decision-making systems \cite{DengYue2017}.

In this sense, several \textit{Reinforcement Learning} (RL) works \cite{MoodyJohn1997, DengYue2017, Aboussalah2020, Ye2020} 
have demonstrated benefits when solving active trading tasks through a decision-making process that learns the best strategy to maximize long-term profitability on a given asset. When designing RL systems, researchers typically adopt feature representation that relates exclusively to the time series of asset prices \cite{MoodyJohn1997, DengYue2017, Aboussalah2020}.
Some approaches to ML in finance \cite{KhadjehNassirtoussi2014, Henrique2019} show that combining features derived from price time series with textual information from external financial news provides good results in the market forecasting task based on supervised learning.

Notably, some RL authors \cite{Feuerriegel2016, YangSteveY2018, Ye2020} have recently successfully explored natural language processing techniques to extract features for a sequential decision-making process.
Despite the successful efforts in this promising new path, many questions still need to be investigated, in particular the \textit{Market sentiment momentum} and the \textit{Instability of RL methods}. The first concerns the extraction and capture of the prevailing mood of the market on a given asset \cite{KhadjehNassirtoussi2014}, and the latter refers to the instability and difficulty of generalizing of RL techniques \cite{Henderson2018} -- problems amplified by the stochasticity of the financial market environment \cite{Tsay2010}.

We approach these issues by proposing the \textit{Sentiment-Aware RL intelligent trading system} (SentARL). This architecture displays a modular design that facilitates incorporating past periods sentiment features in an adjustable form that accounts for the persistent market mood and the dissipation of news impact over time. 
Furthermore, this design handles the \textit{textual data coverage} \cite{Ye2020} from the absence of news articles at every instant. 
An important novelty of this design is that we are the first RL architecture to benefit from using a state-of-the-art sentiment extractor that takes inspiration from the optimal configuration \cite{LimaPaiva2020} of the winner architecture \cite{Mansar2017} of the SemEval-2017 Task 5 competition \cite{Cortis2017}. Also, this competition provided a gold-standard dataset of textual news headlines labeled by market specialists, which we use to train our sentiment extractor.

Considering the research issues we are approaching, we adopt a methodology that embraces suggestions given by \cite{Henderson2018} about proper experimental design and evaluation of RL methods. 
Thus, to thoroughly evaluate SentARL, we gathered data for twenty different assets from various segments and employed a rolling window evaluation setup that tests five periods, including the COVID-19 pandemic crisis. 
For market transaction costs (TC), we employ both a no-penalty and a high-penalty environment.

Our ablation study compared SentARL with its sentiment-free baseline version. Next, even though not all prominent RL research adopts the classic buy-and-hold (BH) strategy as a baseline, we will indeed use it as so. In essence, using the BH strategy as a baseline can help investigate our system's viability in the real world.
Moreover, we ran five trials with different model seed initializations for each combination of asset, period, and TC, totaling a thousand trial combinations for comparison between approaches.
Running such a high number of trials makes performance analysis more robust by helping establish how steadily a system can outperform the baselines and, thus, prevents sporadic occurrences from being taken as the norm.
For instance, rarely have previous research performed such an extensive examination of a proposed model under so many circumstances such as the ones here presented. Hence, the current work examines how difficult it is to produce a system that reliably outperforms the baselines under different scenarios.
Finally, valuation metrics included accumulated total return and Sharpe ratio \cite{Sharpe1966}. 

Our contributions here are twofold: 
\begin{itemize}
    \item \textbf{SentARL}, a new Sentiment-Aware RL intelligent trading system for improving the profitability of asset trading in an RL sequential decision process by efficiently leveraging market sentiment momentum;
    \item First RL active asset trading approach that not only incorporates sentiment information from news articles but also adopts a methodology that follows recent guidelines for an experimental design suitable for evaluating the stability, generalization, and consistency of the RL methods \cite{Henderson2018}.
\end{itemize}

Our results show that SentARL outperforms the sentiment-free baseline by 35\% in average total return (TR) while exhibiting improved stability for 13 assets (given minimal coverage and correlation), TCs, initializations, and periods (including the high volatility COVID-19 Pandemic). Moreover, SentARL consistently outperforms the BH strategy for 11 assets achieving a 17\% higher average TR in the no-penalty environment. We also concluded that:
    \begin{itemize}
        \item There is a lower bound requirement to the textual data coverage and correlation that favors better results and may guide the selection of assets.
        \item Market mood information helps shape the system strategy and accounts for current and future actions.
    \end{itemize}

\section{Related Work}
\label{sec:related-work}

When devising good trading strategies, RL methods prove to be a natural solution, given their learning policies that map observed situations to available trades aiming for high rewards.
The first active trading RL article dates back to 1997 and gained notoriety by introducing the \textit{Recurrent RL} (RRL) architecture with a variation of the \textit{policy gradient} method for direct approximation of parametric policies from gradient ascent over past actions and rewards \cite{MoodyJohn1997}. 
This pioneering work was so influential that researchers are still extending its basic structure and ideas \cite{DengYue2017, Aboussalah2020}. In our work, we adopt two original RRL concepts. First, we have the concise formulation for calculating the financial return, given the continuous change in trading positions, price differences, number of invested shares, and TC. Second, we have the state representation that reflects the agent's interaction with the market by combining historical prices and past action.

RRL \cite{MoodyJohn1997} falls into the \textit{policy-based} RL category, which exhibits convergence to local optima and high variance issues. Alternatively, \textit{value-based} methods such as Deep Q-Learning (DQN) \cite{Mnih2015} present lower variance but suffer from bias. Later, the \textit{actor-critic} methods emerge as a hybrid attempt to address the weaknesses mentioned earlier. For instance, the Asynchronous Advantage Actor-Critic (A3C) algorithm has outperformed previous state-of-the-art DQN methods in the video game domain  \cite{Mnih2016}. 
An active trading problem study promptly replicated this comparison \cite{LiYang2019} and reached similar conclusions about the potential of the A3C architecture. Given these promising results, we adopt the synchronous algorithm \textit{Advantage Actor-Critical} (A2C) \cite {Mnih2016}, less explored but equally effective according to previous work \cite{WuYuhuai2017}.

Extracting market information from textual news for market prediction using supervised learning has long been a promising path \cite{Bollen2011, HuZiniu2018}. Moreover, surveys on ML techniques \cite{KhadjehNassirtoussi2014, Henrique2019} identified a large body of research that adopts textual news information and price time-series for market forecasting. However, only recently a few RL works have investigated this approach. 
Feuerriegel and Prendinger \cite{Feuerriegel2016} used a sentiment dictionary (i.e., a lexicon) to extract word-level sentiment from textual news as a complementary input to its Q-learning algorithm. 
Other work \cite{YangSteveY2018} employed the inverse RL method to produce sentiment-charged reward features that act as proxies to investors' sentiment. Still, instead of using these sentiment reward features for learning a trading strategy, this approach placed them into a supervised market predictor. 
Recently, Ye et al. \cite{Ye2020} adopted deep learning-based methods to generate word embeddings and market predictions to augment the RL market's state representation, but for the portfolio management problem instead of single asset trading.

Similar to previous works \cite{Feuerriegel2016, Ye2020}, we aim to improve the state representation with textual news features. However, unlike Ye et al. \cite{Ye2020}, we stayed away from using news-based market trend predictions as features for our model. Also, aiming at capturing the sentiment market momentum, we extracted explicit sentiment information (i.e., headlines sentiment scores) instead of implicit latent features (i.e., word embeddings) \cite{Ye2020}. Thus, our approach favors the human examination of extracted information's influence over financial return \cite{Glasserman2020} and its impact on the agent's behavior. Furthermore, our sentiment extractor design benefits from a joint effort by the SA and financial communities in the SemEval-2017 Task 5 competition \cite {Cortis2017}. Several researchers competed to design the best architecture at matching these labels. Therefore, we adopt the state-of-the-art winner design \cite{Mansar2017} as an inspiration for our sentiment extractor. In addition, we employ an auspicious configuration \cite{LimaPaiva2020} to improve the feature extractor performance.

\section{Problem Definition}
\label{sec:problem-definition}

Single asset trading is a financial market task that concerns achieving maximum profitability over an asset by adequately deciding on a position to assume (e.g., buy, sell) over a given period. Unlike portfolio management problems where an investor balances the amount invested in each asset, there is isolation between asset operations in single asset trading. Typically, traders devise strategies that work as plans for selecting operations over an asset according to observed market information. Market information can come from technical indicators or fundamental indicators built upon various sources (e.g., asset price, company reports, financial news). The sampling of market information and subsequent decision-making can occur for several frequencies (e.g., hourly, daily, and others). In this work, we design a market environment with hourly information sampling and decision-making capabilities that allow proper trader interaction to learn and test the trading strategies. Typically, market information includes historical bid and ask OHLC (i.e., open, high, low, close) asset prices. However, as a simplification, our market environment adopts only the hourly close bid price of each asset, \(p_t\). We assume that operations are processed instantly at the decision instant \(t\), where \( T > t > 0 \in \mathbb{Z}\), with the maximum available data point \(T\). Subsequently, we denote the consecutive asset price difference as
\begin{equation}
    \label{eq:price_diff}
    z_t = p_t - p_{t-1} \text{.}
\end{equation}

The trader agent has access to past price information, past financial news headlines regarding the target asset, and all its past trades performed. However, even with the increased information from news, taking operations that match predicted market trends can be quite hard \cite{Ye2020} due to the chaotic nature of the market \cite{Tsay2010}. To avoid this market prediction pitfall \cite{DengYue2017}, learned strategies should avoid frequent position shifting by identifying market momentum. Also, the market stochasticity can lead the agent to encounter very different circumstances between training and test environments and incur poor generalization performance. Leveraging market mood -- or sentiment momentum -- can help both stabilize strategies and make them generalize well between training and test sets. Thus, we establish the trading agent's ultimate goal as learning a trading strategy that consistently maximizes the accumulated financial total return (i.e., profit over initial wealth) by interacting with the environment. We use an RL agent for this purpose.

We formalize this single asset trading problem as a Markovian decision process (MDP) given by the tuple \(\langle \mathcal{S, A, T, R}\rangle\). \(\mathcal{S}\) is the set of states (i.e., conditions) of the environment, and \(\mathcal{A}\) is the set of actions (i.e., operations) available to the agent. The state transition function \(\mathcal{T}: \mathcal{S}\times \mathcal{A}\mapsto P(\mathcal{S})\) describes the probability of transitioning from states after a given action and describes the environment dynamics stochasticity. \(\mathcal{R}: \mathcal{S}\times \mathcal{A}\mapsto \R\) is the function of the reward received for taking actions at given states.

An RL agent solves the MDP as follows: the agent observes the current state $S_t \in \mathcal{S}$ and decides to perform an action $a_t \in \mathcal{A}$. The environment transitions to $S_{t+1}$ with the execution of $a_t$ in $S_t$ and the agent receives a reward $R_t$. The process then repeats. The agent's goal is to find a policy $\pi$ that maximizes the future expected reward $G_t$ given by \cite{Sutton2018}
\begin {equation}
    \label{eq:acc_reward}
    G_t = R_{t+1} + \gamma R_{t+2} + \gamma^2 R_{t+3} + \cdots,
\end {equation}
where $\gamma$ is the discount factor that makes a compromise between immediate and later rewards.

In this environment, the available agent's decisions are Long (i.e., buying), Short (i.e., borrowing), or staying out of the market (i.e., neutral). Also, these operations happen on a fixed amount of the asset's shares which the agent can not control. Ultimately, the environment assumes that market operations can incur TC that can significantly affect the financial return and penalize the agent for frequent position shifts. Therefore, we consider it worthwhile to investigate the high-penalty and no-penalty scenarios.

\section{Solving the problem: SentARL}
\label{sec:solving}

\begin{figure*}[ht!]
  \centering
  \includegraphics[width=0.70\linewidth]{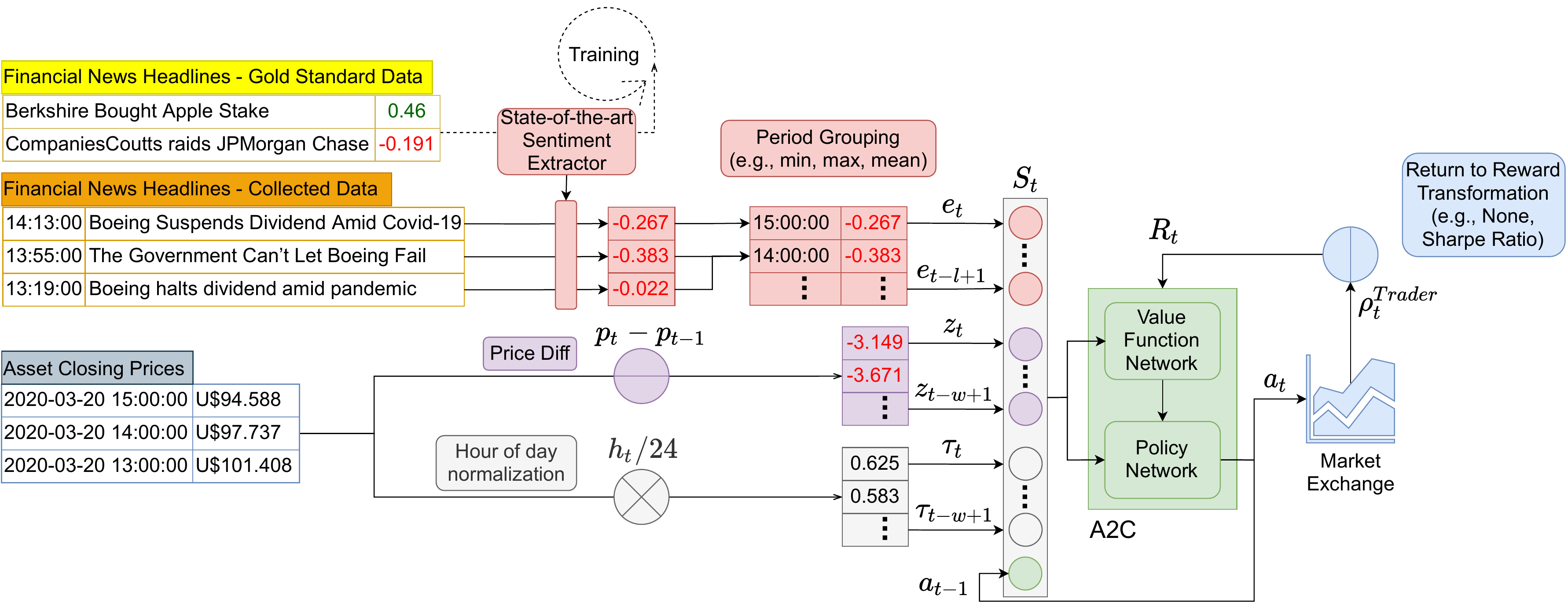}
  \caption{The Sentiment-Aware Reinforcement Learning (SentARL) trading architecture. 
  SentARL employs a news headline sentiment extractor trained with gold standard data; the sentiment scores are then grouped by period to represent the market mood \(e_t\) in the last \(l\) instants. Differences in previous asset's closing prices, \(z_t\) and the normalized time of each data point \(\tau_t\) also compose the market state, along with the last action taken by the agent. Based on \(S_t\), A2C decides \(a_t\) and receives \(R_t\).}
  \Description{In the figure, we have three inputs: the financial news - gold standard data to train the state-of-the-art sentiment extractor, a financial news headline collected data which is used by the sentiment extractor to generate the sentiment signals $e_t,..,e_{t-l+1}$, and we have the asset closing price which we treat by calculating the asset price return, $z_t = p_t-p_{t-1}$ and providing a historical window of size $w$, $z_t,...,z_{t-w+1}$ and the normalized \(\tau_t = h_t/24\) historical hour of the day, $\tau,...,\tau_{t-w+1}$. These three inputs compose the state $S_t$, which the A2c algorithm consumes both by the value function network and the policy network. Also, the A2c algorithm consumes the last action given in the time $t-1$. As output, the A2c algorithm returns the action to be employed in the time $t$. By comparing the action with the next market prices, we have the reward $R_t$, the Sharpe ratio after-action $a_t$.}
  \label{fig:general_architecture}
\end{figure*}

Given the single asset trading formulation as an MDP, an RL algorithm learns a policy (i.e., a strategy) for maximizing trading profits by interacting with the environment on a trial-and-error basis. We propose SentARL as an RL modular architecture for incorporating fundamental indicators from news via an explicit market mood information extraction preprocessing. Fig. \ref{fig:general_architecture} depicts the RL  sentiment-aware agent by introducing past market sentiment features to the agent's state representation. SentARL employs an A2C algorithm with additional features such as past assets' closing prices, hours of trading prices, and last action. This flexible architecture allows experimentation with several sentiment grouping methods (e.g., min, mean, max) and sentiment features' quantity.

We formalize the trading operations available to the agent as the action set \(\mathcal{A} = \{Long, Neutral, Short\} \doteq \{1, 0, -1\}\). At each instant \(t\) the agent selects an action \(a_t = a \in \mathcal{A}\) for trading a fixed amount \(\varphi\) of an asset. Part of the state variable consists of the concatenation (represented by $\mathbin\Vert$) of the price difference time-series with its corresponding normalized hour of the day,
\begin{equation} \label{eq:tech_ind}
    S_t^{M}=[z_t,..., z_{t-w+1}] \mathbin\Vert [\tau_t,..., \tau_{t-w+1}] \in \R^w \text{,}
\end{equation}
with a look-back window of size \(w \in \mathbb{Z}\).
Also, \(\tau_t = h_t / 24\) is the normalized hour of the day, where \(h_t\) is the hour of the day at the instant \(t\) in the 24-hour format. There is a part of the state that the agent can influence, which is its positions over assets, i.e., 
\begin{equation} \label{eq:last_action}
    S^A_t = [a_{t-1}] \text{,}
\end{equation}
is the previous action taken by the agent. This part of the state can help stabilize agent's  position shifting \cite{MoodyJohn1997, DengYue2017}. Finally, we have the base state representation \(S_{t}^{B} \in \mathcal{S}\) as such
\begin{equation} \label{eq:state_base}
    \begin{aligned}
        S_{t}^{B} & = S_t^{M} \mathbin\Vert S_t^A \\
            & = [z_t,...,z_{t-w+1} ] \mathbin\Vert [\tau_t,..., \tau_{t-w+1}] \mathbin\Vert a_{t-1}\text{.}
    \end{aligned}
\end{equation}
Later in this section we expand the state representation with market sentiment features.

Calculating the immediate financial return requires taking the product between the trader's last position \(a_{t-1}\) and the number of an asset's shares \(\varphi\) traded since the last price difference \(z_t\), given by \(\varphi \cdot z_t \cdot a_{t-1}\).
Additionally, the decision to take the long or short position can be made at each step but will subject the agent to a TC (\(c \in \R\)) for each position change.
The TC is usually a percentage of the transaction value and can drastically affect the trader's decision frequency and performance. Ultimately, we write the trader's return over a portfolio with only one asset as
\begin{equation}
\ \rho^{Trader}_{t} = \varphi \left[z_t a_{t-1} - c|a_t - a_{t-1}|\right] \text{,}
\label{eq:trader_return}
\end{equation}
where \(a \in \{-1, 0, 1\}\).

The agent receives a reward after the state transition resulting from the application of \(a_t\) to \(S_t\).
The reward, represented by a numerical value, usually positive (a reward) or negative (a penalization), indirectly has the decision problem's desirable goal.
Also, the reward can have delayed, indirect, and occasional characteristics, which can be handled by expected discounted rewards.
Based on the total reward obtained during the interaction with the environment, the agent updates the policy to maximize the accumulated reward over time. 
In this work, the reward is the financial return of the trader's asset price, $R_t = \rho^{Trader}_{t}$, which is calculated by Eq. \ref{eq:trader_return}.

\subsection{Market's Mood Incorporation}
\label{subsec:state_sent_momentum}

Our goal here is to explicitly capture the market sentiment momentum and introduce it into SentARL's state representation. To do this, we reproduced the winner, state-of-the-art design \cite{Mansar2017} from the SemEval-2017 financial sentiment challenge Task 5 \cite{Cortis2017}. This SA competition produced a gold-standard financial dataset through expert labeling of news texts.
We configured and trained this sentiment component for improved performance over the gold-standard dataset provided in SemEval-2017 \footnote{We refer the reader to \cite{Mansar2017, LimaPaiva2020}  for details on this sentiment extractor word-embedding-based deep Convolutional Neural Network.}.
According to the perceived sentiment, market specialists attributed numeric values between \(-1\) (most pessimistic) and \(1\) (most optimistic) to each headline to create this gold-standard dataset. Thus, the final sentiment label for each headline is the average of all values.

Next, we employ a grouping method (e.g., min, mean, max) over the sentiment scores of all news headlines labeled by the sentiment extractor in a given hour to produce the overall emotion at a given instant, \(e_t = [-1,1] \in \R\). Subsequently, to capture the market mood, we have 
\begin{equation} \label{eq:state_sent_momentum}
    S_t^E = [e_t,..., e_{t-l+1}] \in \R^{l} \text{,}
\end{equation}
as the feature vector that represent the hourly look-back sentiment window of size \(l \in \mathbb{Z}\).

Finally, to attain our complete state representation we concatenate representations from Eq.\ref{eq:tech_ind}, \ref{eq:last_action} and \ref{eq:state_sent_momentum}
\begin{equation} \label{eq:state_final}
    \begin{aligned}
        S_t & = S_t^E \mathbin\Vert S_t^{B} \\
            & = [e_t,..., e_{t-l+1}] \mathbin\Vert [z_t,...,z_{t-w+1} ] \mathbin\Vert [\tau_t,...,\tau_{t-w+1} ] \mathbin\Vert a_{t-1}\text{.}
    \end{aligned}
\end{equation}
We designed our architecture with different look-back windows (\(l\) and \(w\)) for each type of feature for guarantying independence and flexibility of experimental verification.

\subsection{Advantage Actor-Critic}
\label{subsec:a2c}

We solve the MDP using the Advantage Actor-Critic (A2C) RL algorithm, which is the synchronous version of the asynchronous actor-critic \cite{Mnih2016}. 
 A2C learns a policy \(\pi: \mathcal{S} \mapsto \mathcal{A}\) for defining which action \(a_t \in \mathcal{A}\) to apply in state \(S_t \in \mathcal{S}\).

The MDP problem is solved when the RL agent finds, based on its acquired experience, a policy $\pi$ that maximizes the agent's portfolio wealth over time.
A2C uses the advantage function $A(S_t,a_t) = Q(S_t,a_t)-V(S_t)$ that can also be written as $A(S_t,a_t) = R_{t}+\gamma V(S_{t+1})-V(S_t)$ \cite{Mnih2016}.

To approximate the value function, we use an artificial neural network (ANN) with parameters $\vartheta_i$ that is updated using the loss $L_i$, in the iteration $i$, calculated by the equation 
\begin{equation}
   \ L_{i}(\vartheta_{i})=\mathbb{E}\left[R_t+\gamma V\left(S_{t+1}; \vartheta_{i-1}\right)-V\left(S_t;\vartheta_{i}\right)\right]^{2},
\end{equation}
which is the \textit{critic} loss part of the A2C algorithm.
Then for the \textit{actor} part, we approximate the policy function \(\pi(a_t \mid S_t; \theta_{i})\) via a second ANN in which parameters $\theta$ are updated by employing the scaled (by the advantage) log-likelihood,
\begin{equation}
   \ \theta_{i+1} \leftarrow \theta_{i} + \alpha \cdot A(S_t,a_t) \nabla \ln \pi(a_t \mid S_t; \theta_{i}) \text{,}
\end{equation}
where $\alpha$ is the learning rate.
Ultimately, by learning both value and policy functions, the A2C approach reduces the bias from a typical critic method and the variance from a conventional actor approach, bringing practical improvements \cite{Mnih2016}.

\section{Experimental Verification}
\label{sec:experiments}

We evaluated our proposal according to two main aspects.
We measure profit-related metrics such as Sharpe ratio (SR), total return (TR), and annualized return (AR), comparing SentARL to the sentiment-free baseline and the BH benchmark in an ablation study. 
In addition, we present an analysis about the total news available for a particular asset and the performance increase of SentARL. Also, for easier reproducibility of results, the market simulation and SentARL implementation is publicly available \footnote{https://github.com/xicocaio/its-sentarl}.

\subsection{Datasets}
\label{subsec:datasets}

To validate our research hypothesis, we selected assets considering aspects such as price, traded volume, and popularity (i.e., higher presence in news articles). Typically, single asset trading works employ less than ten stocks in their simulations \cite{Henrique2019} and, thus, we assume that selecting twenty assets leads to a much more robust verification. Additionally, we diversified our data by selecting assets from companies in different market sectors such as high-tech (AAPL, AMZN, FB, GOOGL, INTC, MSFT, NFLX), financial (JPM, MA, V), consumer discretionary (DIS, HD, JNJ, KO, PFE, PG), energy (XOM), ETF (SPY), industrial (BA) and Communication Services (T). Employed datasets include 5,267 hourly price data points between 2018 to 2020 (three years) collected from a specialized financial portal \footnote{https://www.dukascopy.com/}. Also, all collected news data came from a portal \footnote{https://www.marketwatch.com/} that includes articles from several news outlets (e.g., Wall Street Journal).

Although some companies can be quite popular, the news used for SA only matches some of the data points in the price time series, meaning not all data points in the price time series display a corresponding news release. There is therefore a news coverage issue. For our selection of assets, we have a maximum news count of 5,041 for the ETF of SP500, which means that we can associate 95.71\% of them with prices.
On average, we have 25\% of news coverage, which means that we have at least one news headline for 25\% of the hourly price data points to extract sentiment. The minimum value of coverage is 5.6\% for asset MA. The degree of coverage assesses the assumption that, with more news, we can better capture the sentiment that will affect each price, improving the state representation. We evaluate this assumption in our experiments and show the influence of news on SentARL performance. In addition, we further discuss which assets have the most coverage and how this influences our approach. A notable feature of the data is that news articles primarily discuss past events rather than predicting future trends.

\subsection{Sentiment Correlation Signal}
\label{subsubsec:corr_pulse}

\begin{figure}[htbp]
  \centering
  \includegraphics[width=0.8\linewidth]{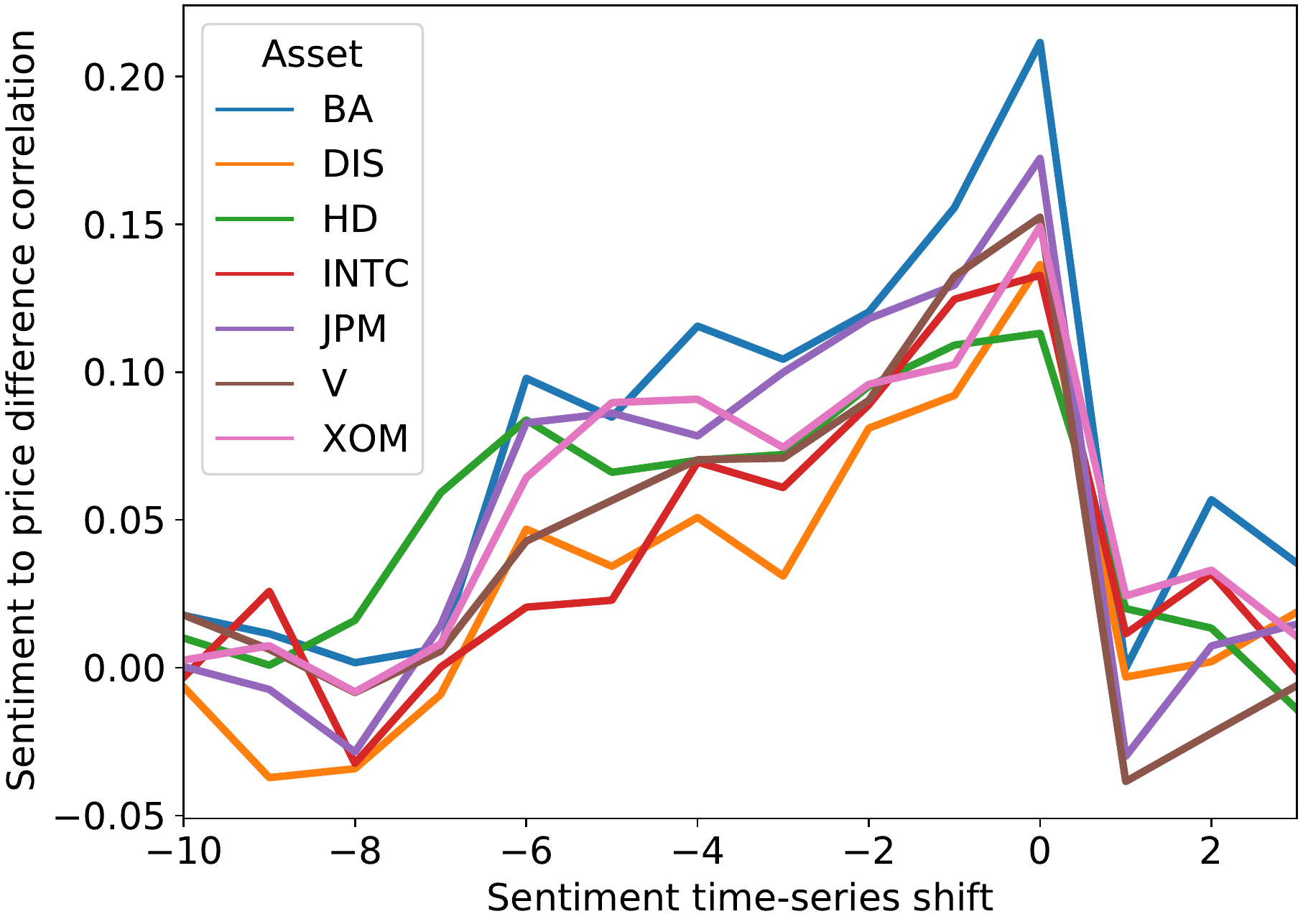}
  \caption{In the x-axis we move the sentiment time-series \(e_t\) in relation to the price \(z_t\) and in the y-axis we calculate the Pearson correlation between these shifted vectors. The rising correlation from the negative shift to zero shows that news sentiment is mostly about past events, and there is poor correlation to the immediate next step (shift + 1).}
  \Description{In the x-axis we move the sentiment time-series \(e_t\) in relation to the price \(z_t\) and in the y-axis we calculate the Pearson correlation between these shifted vectors. The rising correlation from the negative shift to zero shows that news sentiment is mostly about past events, and there is poor correlation to the immediate next step (shift + 1). The y-axis ranges from -0.05 to 0.20 and the x-axis ranges from -10 to 3. The graph depicts seven lines, one for each asset. All the lines have a upward trend until the value of 0 in x-axis. From 0 we have a sharp downward trend changing to a upward trend on 1.}
  \label{fig:corr_pulse}
\end{figure}

We analyze the input of sentiment features for the agent state representation. First, we took the linear Pearson correlation between the time-series of the all grouped hourly sentiment \([e_{t=1},..., e_{t=T}]\) and all price differences \([z_{t=1},..., z_{t=T}]\) of each asset, where \(T\) is the maximum available data point. 
Furthermore, by shifting the sentiment time-series against the price difference and then taking the correlation of some assets, we observe a sentiment correlation pulse in Fig. \ref{fig:corr_pulse}. 
The increase in correlation starts around a shift of \(-7\), where we compare the sentiment score \(e_t\) at instant \(t\) to the price difference seven hours before, \(z_{t-7}\). This progression in correlation value continues according to the shift decrease and reaches its maximum at zero before turning back to the correlation observed much earlier at a shift of \(-10\). This worst correlation at \(+1\) shift indicates that the market can react unpredictably after a relevant event. Also, this correlation pulse seems to confirm that we are capturing the sentiment of past events and that trying to find a direct linear predictive power in this series may not be ideal. Thus, this correlation pulse, together with the experimental results we further discuss, seems to corroborate our hypothesis that observing market sentiment momentum is an adequate and promising research direction and serves as a good indicator of the observed performance of SentARL. Finally, this correlation pulse guided us when selecting the sentiment look-back window size \(l\).

\subsection{Experimental setup and parameters}
\label{subsec:exp-params}

\begin{figure}[htbp]
  \centering
  \includegraphics[width=0.7\linewidth]{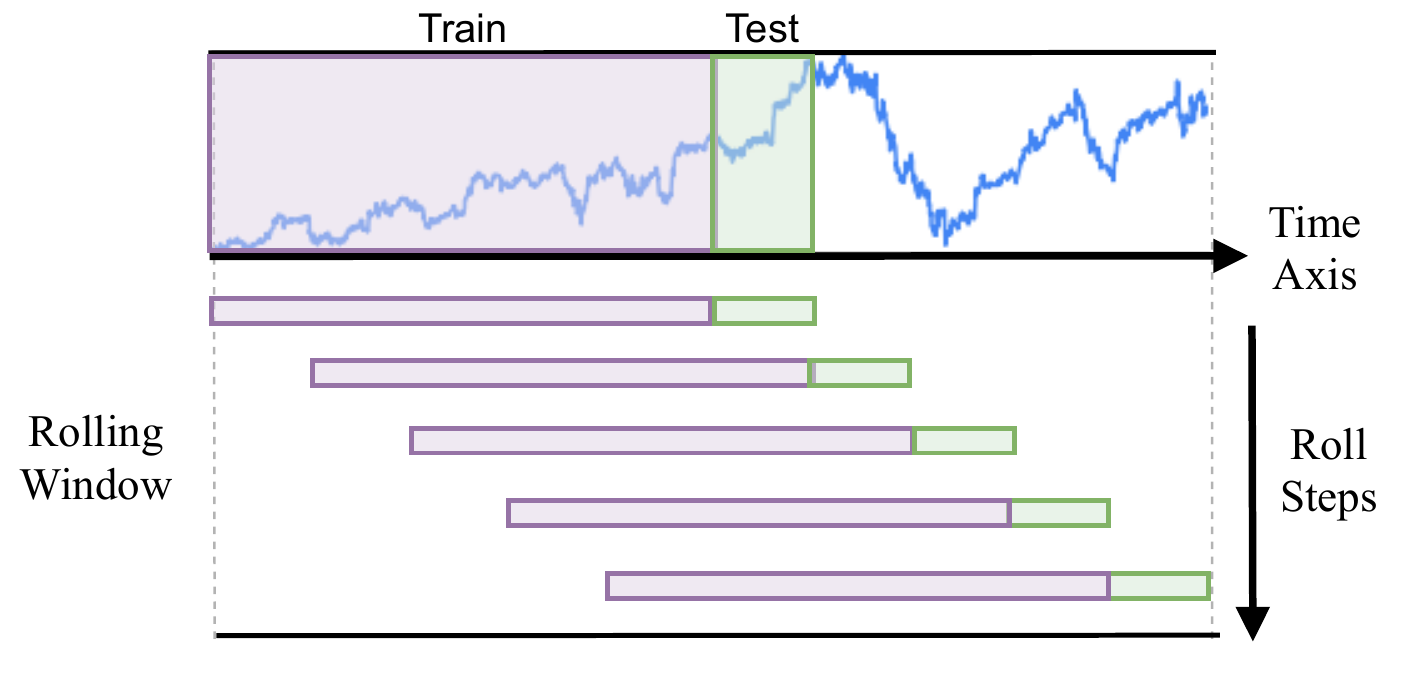}
  \caption{Experimental rolling window setup in relation to splitting the training and testing dataset, with five consecutive forward roll steps in the period.}
  \Description{Experimental rolling window setup in relation to splitting the training and testing dataset, with five consecutive forward roll steps in the period. The graph depicts an price time series in the upper part with an division indicating which part is training and which is test. the test part ends before the end of the time series. The exceeding part divided but the number of rolling windows tells us how much is the displacement of the shifting of each training and test. In the bottom part we have five bars shifted by the displacement indicated describing before, graphically representing each rolling window of training and test.}
  \label{fig:rolling_window}
\end{figure}

We divided the dataset into five equal-sized periods defined by the rolling window in Fig. \ref{fig:rolling_window}. Each window encompasses a training set with 3,377 data points covering two years and a testing set with 374 data points covering around 77 consecutive days (train-test split at an approximate ratio of 0.9 to 0.1), with a stride of also 374 data points. This setup allows evaluating models without look-ahead issues or superposition of testing sets. Ultimately, the rolling window approach helps to investigate performance under different market periods with varying circumstances. For instance, the first window (from 2019-12-09 to 2020-02-26) covers a period of market growth and relative stability, while the second (from 2020-02-27 to 2020-05-13) includes a market crash (Covid-19 pandemic).

In single asset trading research, it is usual to define a fixed amount of shares \(\varphi\) the agent can operate over \cite{MoodyJohn1997, DengYue2017}. We follow this suggestion and define this fixed amount to simplify results analysis as \(\varphi = 1\).
We performed experiments using two different TC: no penalty (0.0\%) and high penalty (0.25\%).
Furthermore, ANNs' weight initialization (given by the seed value) dramatically affects RL results \cite{Henderson2018}. Moreover, the intrinsically noisy price data amplifies this variability problem. Thus, we employed five weights initialization on each training set for better assessing overall performance.
To train the RL agent we consider each interaction with whole price time series an episode. We found that after 100 episodes there is no significant change in the testing set performance.

To define the final SentARL configuration, we executed hyperparameter selection in training using 10\% of the training set for validation in a high-penalty environment with fewer assets and initializations. 
Performing an exhaustive selection would have been prohibitively time-consuming, given the high number of executed trials. Consequently, further investigation should be performed to determine the exact best configuration setup for SentARL.
The selected hyperparameters were:

\begin{description}
    \item [Sentiment grouping method:] Choosing the minimum sentiment score of all headlines in an hour worked slightly better than taking the average or the maximum value.
    \item [Look-back window:] A sentiment look-back window \(l = 5\) worked better than other values (1, 3 ,10). The price and hour-of-the-day look-back windows seemed to be limiting the importance of sentiment features, and the value of \(w = 20\) led to better results. Given the sentiment correlation pulse previously discussed, we should investigate other look-back windows in the future.
    \item [A2C policy and value networks:] We adopted separate networks for the policy and value functions, albeit with similar configurations. Both use a deep ANN multilayer perceptron with 2 hidden layers with 64 nodes each. Larger networks seemed to produce worse results. The learning rate used was 0.99.
    \item [A2C update steps:] A2C performs batch updates to the ANNs after a number of steps, and we found out that 5 steps was a reasonable value.
\end{description}

\subsection{Evaluation metrics and baselines}
\label{subsec:evalmetric}

We report standard metrics for measuring the performance of autonomous trading agents.
Initially, we have the TR given by the accumulated return from each instant (Eq. \ref{eq:trader_return}) over the initial wealth \(\psi\) as formulated below
\begin{equation} \label{eq:total_profit}
    \begin{aligned}
        TR = \frac{\sum_{t=1}^{T} \rho^{Trader}_{t}}{\psi} \text{.}
    \end{aligned}
\end{equation}
The initial wealth is equivalent to the initial amount of cash required to buy \(\varphi\) shares of the given asset at a price \(p^{asset}_{1}\) at the instant \(t=1\) (i.e., \(\psi = \varphi \cdot p^{asset}_{1}\)).
Then, a metric such as the annualized return (AR) is recommended to compare research with data with different periods and sizes. AR is given by
\begin{equation}
    \label{eq:anualized_return}
    AR = (1 + TR)^\frac{D}{365} - 1  \text{,}
\end{equation}
where $TR$ regards the tested period, and $D$ is the total number of trading days on the test set (77 days).
Ultimately, the $TR$ for all trials of a model $k$ -- meaning the different initializations and periods -- determine the SR metric, formulated with inspiration from previous work \cite{MoodyJohn1997, DengYue2017} as
\begin{equation}
    \label{eq:sharpe_ratio}
    SR = \frac{avg(TR_k)}{std(TR_k)} \text{.}
\end{equation}

We compared SentARL with a sentiment-free version (ablation study) and the BH strategy benchmark.
BH consists of buying the asset and holding it during the designated period. This strategy reflects market trends and has the advantage of no TC since there is no transaction. On the other hand, BH's disadvantage is that this benchmark will not capture profit opportunities resulting from market patterns.

\subsection{Results and Discussion}
\label{subsec:results-discussion}

Here we investigate the performance of SentARL and other baselines to verify proper addressing of issues mentioned earlier: market sentiment momentum and instability of RL methods.

% \noindent
% \vspace{0.2}
% \textbf{Model generalization.}
\subsubsection{Model generalization.}
\begin{figure}[htbp]
  \centering
  \includegraphics[width=\linewidth]{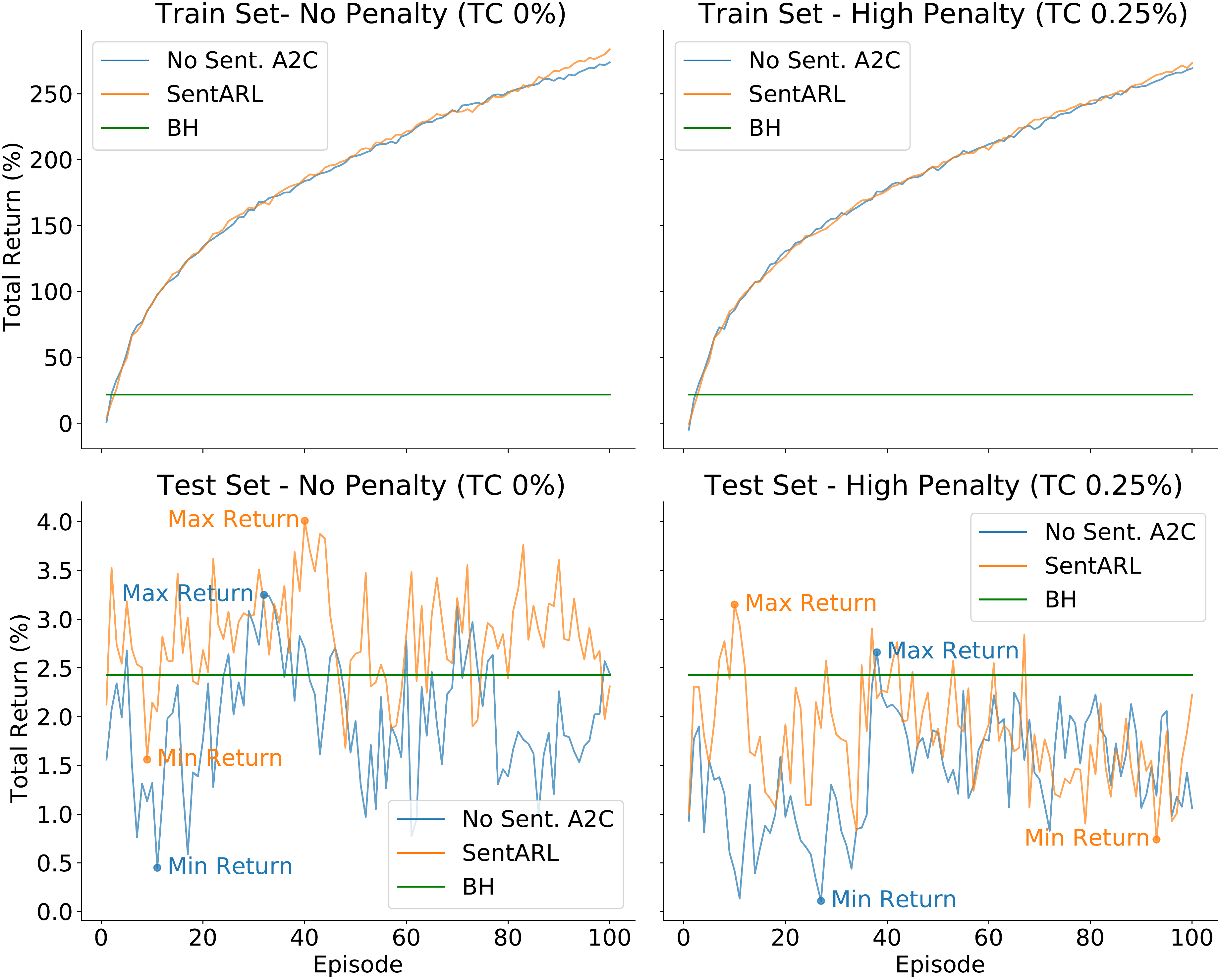}
  \caption{Comparison of SentARL and sentiment-free model performance in the training and test sets according to average TR overall assets by episode, with  BH as reference. Top: Training Set; Bottom: Test Set; Left: No-penalty scenario (TC 0.0\%); Right: High-penalty (TC 0.25\%).}
  \Description{These figures show line plots wherein on the x-axis we have the current train episode, and on the y-axis, we have the average TR overall assets. We have the training set in the top plots, and at the bottom plots, we have the test set. The left plots are the no penalty scenario, and the right ones are the high penalty. The BH as a straight line serves as a reference. Here we observe that SentARL is more consistent than its sentiment-free counterpart as it presents better overall results over most training episodes. We also observe that SentARL consistently outperforms the BH in the test no-penalty scenario, while in high-penalty, it occurs more rarely.}
  \label{fig:val_episode_curve}
\end{figure}

Fig. \ref{fig:val_episode_curve} shows the average TR of all assets per episode in training and per test sets for both training costs (TC 0\% and TC 0.25\%) -- for SentARL, for the sentiment-free version of SentARL (No Sent. A2C in Figure) and the average TR for baseline strategy BH (constant line). 
Here we observe that the training profiles for SentARL and the sentiment-free baseline are pretty similar for both TCs. They increase steadily and indefinitely, reaching ten times greater average TR than the BH baseline. 
We expected such an outcome since we are dealing with an environment with chaotic characteristics \cite{Tsay2010}. However, when looking at the evolution of the average TR by episode in the test set, both SentARL and the sentiment-free baseline exhibit erratic behavior, making it hard to define when overfitting may have started. In this sense, we confirm that traditional machine learning approaches for overfitting verification may not be sufficient to determine the generalization capacity in trading environments.

\subsubsection{Overall evaluation.}
Although, according to Fig. \ref{fig:val_episode_curve}, generalization is a challenge and the performance of both models fluctuates considerably, SentARL seems to have learned a very profitable trading strategy for the no-penalty scenario: SentARL outperformed the sentiment-free model in 89 of the 100 episodes with a 43\% higher average TR. Moreover, SentARL outperformed the BH strategy in 82 episodes with a 17\% higher average TR. Notably, when we introduce a high TC, there is a decrease in the performance of both models (SentARL outperforms BH in only 14 episodes). Still, SentARL exhibited 27\% better performance than the sentiment-free baseline and attained better returns in 66 episodes. Interestingly, we observe that SentARL performance is better in the initial 50 episodes, while for the sentiment-free model, it is the opposite. Looking at Table \ref{tab:overall_results} we observe more results. SentARL is the only strategy that outperforms the BH strategy in average TR (16.5\% higher), but only when there are no costs involved.

\begin{table}[h]
  \caption{Overall results of SentARL, the baseline and BH according to TR, AR and SR.}
  \label{tab:overall_results}
  \begin{tabular}{llllll}
    \toprule
    TC (\%) & Strategy         & Mean TR & Mean AR & SR \\
    \midrule
    - & BH    & \textbf{2.43\%}   & \textbf{12.03\%}   & \textbf{0.64} \\
    0.0  & No Sent.  & 1.98\%  & 9.73\% &  0.44 \\
            & SentARL   & \underline{2.83\%}  & \underline{14.12\%} &  \underline{0.62} \\
    0.25 & No Sent. & 1.41\%  & 6.86\% & 0.19  \\
            & SentARL & \underline{1.79\%}  & \underline{8.8\%} &  \underline{0.51} \\
  \bottomrule
\end{tabular}
\end{table}

\subsubsection{Algorithms' Consistency.}
We observe in Fig. \ref{fig:val_episode_curve} and Table \ref{tab:overall_results} that for both transaction cost scenarios, SentARL achieved higher episode maximums and minimums, observed less difference between maximum and minimum episode values, and achieved higher SR. Particularly for the high-penalty scenario, SentARL achieved its maximum average TR of 3.15\% at episode 10, which is 18\% greater than the sentiment-free baseline maximum average TR of 2.66\% achieved at episode 38. SentARL achieved its minimum average TR of 0.74\% at episode 93, 592\% greater than the sentiment-free baseline minimum average TR of 0.11\% achieved at episode 27. Also, SentARL reached a SR of 0.51 in the high-penalty scenario versus the 0.19 value of the sentiment-free baseline. In conclusion, SentARL showed evidence that incorporating market mood can improve the stability of learned strategies.

\subsubsection{Results by asset.}
In Table \ref{tab:sr_asset} we compare SentARL with BH and the sentiment-free approach according to the SR by asset and TC. Although these results already show the capability of SentARL to leverage market mood, we will look deeper down to investigate the details about its performance, benefits, and when it might work better in the following section. Ultimately, observe that for certain assets (e.g., BA, INTC, XOM) SentARL can outperform the BH.

\begin{table}[ht]
    \caption{Sharpe ratio by asset and TC for SentARL, sentiment-free baseline (No Sent.) and BH. Underlined values identify the best performance between models for a given transaction cost. Bold values identify best performance considering high penalty scenario and BH.}
    \label{tab:sr_asset}
    \begin{tabular}{lrrrrr}
    \toprule
    & \multicolumn{1}{l}{} & \multicolumn{2}{c}{TC 0.0\%}  & \multicolumn{2}{c}{TC 0.25\%}                          \\
    Asset & BH & SentARL & No Sen. & SentARL & No Sen.\\
    \midrule
AAPL  & -0.125          & 0.423        & {\underline{ 1.439}}  & 0.082                 & {\underline{ \textbf{1.299}}}  \\
AMZN  & \textbf{1.120}  & {\underline{ 0.497}}  & 0.263        & {\underline{ 0.454}}           & 0.190                 \\
BA    & -0.056          & {\underline{ 1.738}}  & -0.136       & {\underline{ \textbf{0.665}}}  & 0.083                 \\
DIS   & \textbf{0.252}  & -0.489       & {\underline{ -0.361}} & {\underline{ -0.291}}          & -0.484                \\
FB    & \textbf{0.704}  & {\underline{ 0.042}}  & -0.025       & -0.102                & {\underline{ 0.156}}           \\
GOOGL & \textbf{0.742}  & -0.280       & {\underline{ 0.226}}  & -0.287                & {\underline{ -0.059}}          \\
HD    & \textbf{0.490}  & {\underline{ 1.273}}  & 1.047        & {\underline{ 0.412}}           & 0.000                 \\
INTC  & -0.205          & {\underline{ -0.175}} & -0.662       & {\underline{ \textbf{0.554}}}  & -0.351                \\
JNJ   & \textbf{0.912}  & {\underline{ 0.529}}  & -0.209       & {\underline{ 0.189}}           & -0.428                \\
JPM   & \textbf{0.037}  & {\underline{ -0.195}} & -0.369       & {\underline{ -0.150}}          & -0.435                \\
KO    & \textbf{0.056}  & -0.192       & {\underline{ -0.171}} & {\underline{ -0.073}}          & -0.109                \\
MA    & 0.414           & {\underline{ 1.556}}  & 0.807        & {\underline{ \textbf{1.379}}}  & 0.762                 \\
MSFT  & \textbf{1.792}  & {\underline{ 1.112}}  & 0.778        & {\underline{ 0.915}}           & 0.629                 \\
NFLX  & \textbf{1.249}  & {\underline{ 0.501}}  & 0.194        & 0.128                 & {\underline{ 0.554}}           \\
PFE   & -0.177          & -0.284       & {\underline{ -0.072}} & -0.159                & {\underline{ \textbf{-0.155}}} \\
PG    & 0.264           & {\underline{ 0.674}}  & 0.523        & {\underline{ \textbf{0.572}}}  & 0.469                 \\
SPY   & 0.384           & {\underline{ 0.940}}  & 0.857        & 0.517                 & {\underline{ \textbf{0.535}}}  \\
T     & \textbf{-0.433} & -0.336       & {\underline{ -0.289}} & {\underline{ -0.472}}          & -0.770                \\
V     & 0.594           & 1.059        & {\underline{ 1.835}}  & {\underline{ \textbf{1.242}}}  & 0.692                 \\
XOM   & -0.455          & {\underline{ 0.009}}  & 0.080        & {\underline{ \textbf{-0.273}}} & -0.555               \\
    \bottomrule
\end{tabular}
\end{table}

\subsubsection{News coverage and Sentiment Correlation thresholds.}
In Fig. \ref{fig:corr_ratio_sentral_vanila_diff}, we investigate the better performance of SentARL over the sentiment-free model by asset depending on news coverage and correlation between measured market sentiment and price time-series for the high-penalty scenario. Initially, looking at the correlation axis (horizontal axis), we observe that, as expected, higher correlation leads to the predominance of SentARL advantage. Also, even a correlation as low as 0.1 can tip the scales in favor of SentARL. Next, observing the news coverage axis (vertical axis), we verify that for similar correlations (e.g., XOM and V), the news coverage can increase SentARL advantage. However, the amount of news seems to harm SentARL performance when the correlation is below a particular value (e.g., AAPL, FB, GOOGL, NFLX). Interestingly, in the case of AMZN, this does not occur and could be an outlier. Moreover, we notice that the SPY may be a problematic asset to capture sentiment, as it is composed of the price of 500 assets, and each news item may only affect a small part of the index.

\begin{figure}[htbp]
  \centering
  \includegraphics[width=0.9\linewidth]{figs/corr_ratio_sentarl_vanila_diff.pdf}
  \caption{Difference in TR (\%) between SentARL and sentiment-free baseline by asset according to news coverage and sentiment to price time-series correlation in the test set with TC 0.25\%. Orange dots indicate situations where SentARL outperformed the sentiment-free base model, while blue dots are the opposite. The intensity of colors indicates the magnitude of the percentage difference between models.}
  \Description{This figure shows a scatter plot with the news coverage ratio in the y-axis, the correlation between measured sentiment, and the close price time-series difference in the x-axis. Orange dots indicate situations where SentARL outperformed the sentiment-free base model, while blue dots are the opposite (e.g., AAPL, FB, GOOGL, MA, NFLX). The intensity of colors indicates the magnitude of the percentual difference between models.}
  \label{fig:corr_ratio_sentral_vanila_diff}
\end{figure}

It is notable that some of the assets with the worst correlation that led to the worst SentARL performance are mainly from the tech segment. It seems to indicate that news regarding tech companies, even though more popular and common, could be more speculative and lead to a false perception of the prevalent market mood. 
Overall, we see a potential threshold above 11\% news coverage and a correlation of 0.128, where some of SentARL's highest performance gains over its sentiment-free counterpart occur, and the chances of getting better SentARL results are possibly higher. Nonetheless, these conjectures still require further theoretical grounding or robustness checks with more assets so that a stronger general rule regarding these characteristics can be drawn for the rest of the market.

\section{Conclusion and future work}
\label{sec:conclusions}

This paper presented SentARL for the single asset trading task, an effective architecture in identifying market sentiment momentum to achieve higher profit consistency than the sentiment-free baseline. Our experiments show that SentARL outperformed the sentiment-free baseline -- in all metrics -- for both transactions costs. Also, when there are no penalties associated with the transactions, SentARL outperformed the BH strategy. On the other hand, our exhaustive investigation showed how hard it is to consistently outperform the BH in a high-penalty scenario and, thus, prevented us from incorrectly assuming that SentARL had achieved it. Moreover, when compared to the sentiment-free model, our proposed architecture had reduced performance variation considering all TCs values, model parameters initialization, and periods. Finally, we identified a hypothetical lower bound requirement for the textual data coverage and correlation necessary for SentARL to benefit from news information. In this sense, monitoring the current news coverage and sentiment correlation on new data would be necessary to turn this architecture into a live application.

Although we observed assuring results, there are still plenty of promising opportunities for examining the market sentiment momentum incorporation into RL methods. For instance, the NLP community has long adopted social networks as textual sources for trading as it provides a vast amount of data and can bring fresher information. Thus, even though this data source might entail additional data preprocessing to reduce noise, it might support increased trading frequency, better news coverage, and sentiment correlation. Ultimately, further verification is necessary to identify the ideal hyperparameters and state features (e.g., sentiment and price window sizes) configuration for improving results so that SentARL can outperform the BH strategy in a high-penalty scenario.

\begin{acks}

This work was financed in part by \textit{Ita\'{u} Unibanco S.A.} through the \textit{Programa de Bolsas Ita\'{u}} (PBI) of the \textit{Centro de Ci\^{e}ncia de Dados} (C$^2$D, EP-USP), %of \textit{Escola Polit\'{e}cnica} of USP, 
by the \textit{Coordenação de Aperfeiçoamento de Pessoal de Nível Superior} (CAPES Finance Code 001), \textit{Conselho Nacional de Desenvolvimento Científico e Tecnológico} (CNPq  grant 310085/2020-9 and 88882.333380/2019-01),
and by the \textit{Center for Artificial Intelligence} (C4AI-USP), with support from %\textit{Fundação de Amparo à Pesquisa do Estado de São Paulo} (
FAPESP (grant 2019/07665-4) and \textit{IBM Corporation}.
Any views and opinions expressed in this article are those of the authors and do not necessarily reflect the official policy or position of the funding companies.
\end{acks}

\bibliographystyle{ACM-Reference-Format}
\bibliography{references}

\end{document}